\documentclass[11pt,onecolumn]{IEEEtran}
\usepackage{mathpazo}

\usepackage{amsmath}
\usepackage{amsfonts}
\usepackage{latexsym}
\usepackage{amssymb}

\usepackage{upref}
\usepackage{theorem}
\usepackage{graphicx}	
\usepackage{psfrag}
\usepackage{cite}
\usepackage{color}

 \newtheorem{prop}{Proposition$\!$}

 \newtheorem{xmpl}{Example$\!$}

\newtheorem{cnstr}{Construction$\!$}

\newcounter{enumrom}
\renewcommand{\theenumrom}{(\roman{enumrom})}

\makeatletter
\renewcommand{\@endtheorem}{\endtrivlist}
\makeatother

\makeatletter
\renewcommand{\thefigure}{{\@arabic\c@figure}}
\renewcommand{\fnum@figure}{{\bf Figure\,\thefigure}}
\makeatother

\newcommand{\cB}{\mathcal{B}}
\newcommand{\cC}{\mathcal{C}}

\newcommand{\cM}{\mathcal{M}}

\newcommand{\cS}{\mathcal{S}}

\newcommand{\be}[1]{\begin{equation}\label{#1}}
\newcommand{\ee}{\end{equation}}

\renewcommand{\leq}{\leqslant}
\renewcommand{\ge}{\geqslant}
\renewcommand{\geq}{\geqslant}

\newcommand{\Cref}[1]{Co\-ro\-lla\-ry\,\ref{#1}}

\DeclareMathOperator{\perm}{perm}

\DeclareMathOperator{\rank}{rank}

\outer\def\proclaim #1. #2\par{\medbreak
 \noindent{\bf#1.\enspace}{\sl#2\par}%
 \ifdim\lastskip<\medskipamount \removelastskip\penalty55\medskip\fi}

\newtheorem{theo}{Theorem}
\newtheorem{defin}{Definition}
\newtheorem{lem}{Lemma}
\newtheorem{cor}{Corollary}

\begin{document}

\title{\textbf{Optimal Locally Repairable Codes and Connections to Matroid Theory}
\vspace*{-0.2ex}}

  \author{\IEEEauthorblockN{Itzhak Tamo\IEEEauthorrefmark{1}, Dimitris S. Papailiopoulos\IEEEauthorrefmark{3} and Alexandros G. Dimakis\IEEEauthorrefmark{3}
 \thanks{
The material in this paper was presented in part at the
IEEE International Symposium on Information
Theory (ISIT 2013), Istanbul, Turkey, July 2013.} 
  }
\\ \IEEEauthorblockA{\IEEEauthorrefmark{1}Dept. of ECE and Inst. for Systems Research University of Maryland, USA}\\
 \IEEEauthorblockA{\IEEEauthorrefmark{3}Electrical and Computer Engineering University of Texas at Austin, USA}\\
 \texttt{tamo@umd.edu}, \texttt{dimitris@utexas.edu}, \texttt{dimakis@austin.utexas.edu}}


%

\maketitle

\begin{abstract}
Petabyte-scale distributed storage systems are currently transitioning to erasure codes to achieve higher storage efficiency. Classical codes like Reed-Solomon are highly sub-optimal for distributed environments due to their high overhead in single-failure events. 
{\it Locally Repairable Codes} (LRCs) form a new family of codes that are repair efficient. 
In particular, LRCs minimize the number of nodes participating in single node repairs during which they generate small network traffic.
Two large-scale distributed storage systems have already implemented different types of LRCs: Windows Azure Storage and the Hadoop Distributed File System RAID used by Facebook. 
The fundamental bounds for LRCs, namely the best possible distance for a given code locality, were recently discovered, but few explicit constructions exist. 
In this work, we present an explicit and optimal LRCs that are simple to construct.
{\color{black}Our construction  is based on grouping Reed-Solomon (RS) coded symbols to obtain RS coded symbols over a larger finite field.
We then partition these RS symbols in small groups, and re-encode them using a simple local code that offers low repair locality.}
For the analysis of the optimality of the code, we derive a new result on the matroid represented by the code's generator matrix. 
\end{abstract}

\section{Introduction}

Traditional architectures of large-scale storage systems rely on distributed file systems that provide reliability through block replication. Typically, data is split into blocks and three copies of each block are stored
in different storage nodes. The major disadvantage of triple replication is the large storage overhead.
As the amount of stored data is growing faster than hardware infrastructure, this becomes a factor of three in the \textit{storage growth rate}, resulting in a major data center cost bottleneck. 

As is well-known, {\it erasure coding} techniques achieve higher data reliability with considerably smaller storage overhead~\cite{weatherspoon2002erasure}. For that reason different codes are being deployed in production storage clusters. Application scenarios where coding techniques are being currently deployed include cloud storage systems like Windows Azure~\cite{huang2012erasure}, big data analytics clusters 
({\it e.g.}, the Facebook Analytics Hadoop cluster~\cite{sathiamoorthy2013xoring}), archival storage systems, and peer-to-peer storage systems like Cleversafe and Wuala.

It is now understood that classical codes (such as Reed-Solomon) are highly suboptimal for distributed storage repairs~\cite{dimakis2010network}.
For example, the Facebook analytics Hadoop cluster discussed in~\cite{sathiamoorthy2013xoring}, deployed Reed-Solomon (RS) encoding for 8\% of the stored data. That portion of the data generated repair traffic approximately equal 
to 20\% of the total network traffic. Therefore, as discussed in~\cite{sathiamoorthy2013xoring}, 
the main bottleneck in increasing code deployment in storage systems is designing new codes that perform 
well for distributed repairs. 

Three major repair cost metrics have been identified in the recent literature: 
{\it i)} the number of bits communicated in the network, {\it i.e.}, the {\it repair-bandwidth} \cite{dimakis2010network, rashmi2011optimal,suh2011exact, tamo2011mds, cadambe2011optimal, papailiopoulos2011repair}, 
{\it ii)} the number of bits read, the {\it disk-I/O} \cite{tamo2011mds,khan2011search}, and  more recently
{\it iii)} the number of nodes that participate in the repair process, also known as, {\it repair locality}. Each of these metrics is more relevant for different systems and their fundamental limits are not completely understood.
In this work, we focus on the  metric of repair locality, one that seems most relevant for single-location high-connectivity storage clusters.

Locality was identified as a good metric independently by Gopalan {\it et al.} \cite{gopalan2011locality}, Oggier {\it et al.} \cite{oggier2011self}, and Papailiopoulos {\it et al.} \cite{papailiopoulos2012simple}.
Consider a code of length $n$ and $k$ information symbols. The $i$-th symbol of the codeword has locality $r_i$ if it can 
be recovered by accessing at most $r_i$ other code symbols. 
A systematic code has \textit{information symbol locality} $r$,
if all the $k$ information symbols have locality $r$.
Similarly, a code has \textit{all-symbol locality} $r$, if all $n$ symbols have locality $r$.
Codes that have good locality properties were initially studied in \cite{han2007reliable} and \cite{huang2007pyramid}.

In \cite{gopalan2011locality}, a trade-off between code distance, {\it i.e.}, reliability, and information symbol locality, was derived for scalar linear codes. 
In \cite{papailiopoulos2012locally}, an information theoretic trade-off for any code (linear/nonlinear) was derived when considering all symbol locality.
An $(n,k)$ code with (information symbol or all-symbol)  locality $r$ has minimum distance $d$ that is bounded by
\begin{equation}
d\leq n-k-\left\lceil \frac{k}{r}\right\rceil+2.
\label{eq:gopalan}
\end{equation}
Bounds on the code-distance for a given locality were also derived and generalized  in parallel and subsequent works in \cite{prakash2012optimal, kamath2012codes,rawat2012optimal,forbes2013locality,mazumdar2013local,mazumdar2013update,cadambeupper}.

{\color{black}An $(n,k,r)$  {\it locally repairable code} (LRC) is a code of length $n$, that takes as input $k$ information symbols, such that \emph{any} of its $n$ output coded symbols can be recovered by accessing and processing \emph{at most} $r$ other symbols (i.e., an LRC has all-symbol locality). }
Codes with all-symbol locality that meet the above bound on the distance are termed \emph{optimal LRCs} and are known to exist when $(r+1)$ divides $n$ \cite{gopalan2011locality,papailiopoulos2012locally,kamath2012codes,prakash2012optimal,rawat2012optimal}.
Explicit optimal LRC constructions, for some code parameters, were introduced in \cite{prakash2012optimal,kamath2012codes,papailiopoulos2012locally,rawat2012locality,silberstein2012error,rawat2012optimal}.
Some works extend the designs and theoretic bounds to the case where repair bandwidth and locality are jointly optimized during multiple local failures \cite{kamath2012codes,rawat2012optimal}, and to the case where security issues are addressed \cite{rawat2012optimal}.
The construction of practical LRCs is further motivated by the fact that two major distributed storage systems have already implemented different types of LRCs: Windows Azure Storage \cite{huang2012erasure} and the Hadoop Distributed File System RAID used by Facebook \cite{sathiamoorthy2013xoring}.
Designing LRCs with optimal distance for all code parameters $n,k,r$ that are easy to implement is a new and exciting open problem.

{\bf Our Contribution:}
We introduce a new \emph{explicit} family of optimal $(n,k,r)$ LRCs.
Our construction is optimal for any $(n,k,r)$ such that $r+1$ divides $n$.
Furthermore it requires $O(k\log n)$ bits in the description of each symbol and a main advantage is in the simplicity of its design.
The codes are simple to construct and are based on Reed-Solomon coded blocks that are re-encoded in a way that provides low repair locality.
The main theoretical challenge is in proving that they are optimal,  i.e., that they achieve the distance bound in \eqref{eq:gopalan}, with equality.
This is done by first deriving a formula for the minimum distance of \emph{any} linear code in terms of some parameters of the matroid represented by the generator matrix of the code. 
We believe that this result has its own interest even outside the scope of LRCs.
In our case, this result provides a sufficient condition on the generator matrix of an LRC: if the generator matrix satisfies such condition, then the LRC achieves the bound of \eqref{eq:gopalan}. 
{\color{black}We establish this condition by using some properties of the determinant function and polynomials over finite fields.} 


The most related works to ours are the two parallel and independent studies of \cite{kamath2012codes} and \cite{rawat2012optimal}.
There, optimal LRC constructions for similar range of code parameters are presented.
Although these constructions rely on different tools and designs than the ones presented here, it would be of interest to explore further connections.

%
%



The remainder of the paper is organized as follows. 
In Section~\ref{sec:code}, we present our code construction.
In Section~\ref{sec:matroids}, we establish a precise formula of the minimum distance of a linear code in terms of the matroid represented by the generator matrix.
In Section~\ref{sec:optimality}, we use the established results and algebra of polynomials over finite fields to prove the optimality of the construction.
In Section~\ref{fast_encoding}, we show that the construction has several other important properties. 
In Section~\ref{Generalization}, we generalize the code construction for the case where multiple local erasures can be tolerated.

\section{Code Construction}
\begin{figure}[t]
\centerline{
\includegraphics[width=0.7\columnwidth]{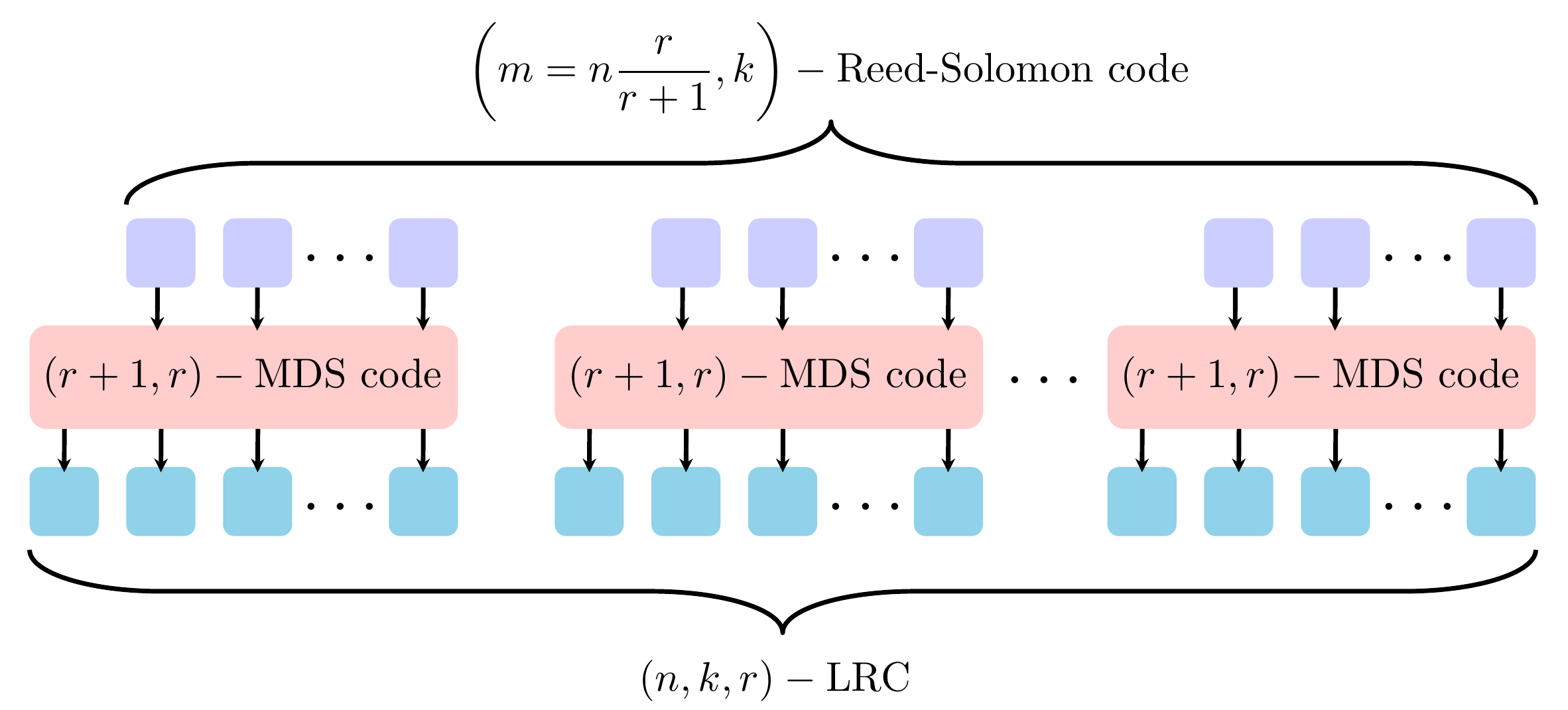}
}
\caption{A sketch of our $(n,k,r)$ LRC construction.
We start with an $(m=n\frac{r}{r+1},k)$-Reed-Solomon code.
We then re-encode the $m$ Reed-Solomon coded symbols in the following manner: divide the $m$ symbols, into $\frac{m}{r}$ groups of $r$ symbols, so that the groups do not overlap.
We then re-encode each group using a specific $(r+1,r)$ MDS code.
The $n$ outputs of the $\frac{m}{r}$ local codes are the encoded symbols of our LRC.
It should not be hard to see that the locality $r$ can be trivially obtained by the local codes: if a symbol is missing the remaining $r$ coded symbols in its group can be used to reconstruct it; 
this is a property of the MDS local code.
Although the code as presented in this figure is not in systematic form, it can be easily done so by a linear transformation applied on the generator matrix.
}
\label{fig:LRC_sketch}
\end{figure}

\label{sec:code}
In this section we present a simple construction of an {\it optimal} $(n,k,r)$ LRC, assuming that  $r+1$ divides $n.$
 Let $m=n\frac{r}{r+1}$ and assume that $1<r<k$.\footnote{If $r=k$, then any $(n,k)$ MDS code is an optimal $(n,k,r=k)$ LRC. Moreover, if $r=1$, then it can be shown that since $r$ divides $k$, $r+1=2$ has to divide $n$, i.e., $n$ is even. 
Replicating each symbol twice in an $(n/2,k)$ MDS code results in an optimal $(n,k,r=1)$ LRC.}

In our construction, we take the output of an $(m,k)$-Reed-Solomon (RS) code, divide it into  $\frac{m}{r}$ non-overlapping groups, each consisting of $r$ coded symbols, then we re-encode the symbols of each group into $r+1$ new symbols using a \emph{specific} $(r+1,r)$-MDS code.
We refer to this $(r+1,r)$-MDS code that we use to re-encode every group of $r$ RS coded symbols, as a {\it local code}.
This construction will be shown to have {\it i)} the desired locality $r$ and {\it ii)} optimal minimum distance.
In Fig.\ref{fig:LRC_sketch}, we give a sketch of the construction.

More formally, let $\mathbb{F}_p$ be a field of size $p\geq m$, and consider a file that is cut into $k$ symbols $$x=[x_1,\ldots, x_k],$$ where each symbol is an element of the field $\mathbb{F}_{p^{k+1}}$.
These $k$ symbols are encoded into $n$ coded symbols $y=[y_1,...,y_n]$:
\begin{equation}
y = x\cdot G\nonumber
\end{equation}
where $G$ is the $k\times n$ generator matrix, with elements over the field $\mathbb{F}_{p^{k+1}}$. 
The construction of $G$ follows. 

\begin{cnstr}
\label{cnstr1}
Let $\alpha_1,...,\alpha_m$ be $m=\frac{nr}{r+1}$ distinct elements of the field $\mathbb{F}_p$, with $p\ge m$, and $\omega$ be a primitive element of the field $\mathbb{F}_{p^{k+1}}$.
Also let $V$ be a $k\times m$ Vandermonde matrix with its $i$-th column being equal to $$\overline{\alpha}_i=(1,\alpha_i,...,\alpha_i^{k-1})^t.$$
Then, the generator matrix of the code is
\begin{equation}
G=V\cdot(I_{m/r}\otimes A),
\label{eq:ththth}
\end{equation}
where $I_s$ is the identity matrix of size $s$ and $A=(a_{i,j})$ is an $r\times (r+1)$ matrix defined as follows: it has $1$s on the main diagonal, and $\omega$s on the diagonal whose elements $a_{i,j}$ satisfy $j-i=1$.
\end{cnstr}
An example of an $(r=3,r+1=4)$ $A$ matrix is given bellow
\begin{equation}
A= \left[ 
\begin{array}{cccc}
1& \omega & 0 & 0 \\
0& 1 & \omega & 0 \\
0& 0 & 1 & \omega \\
 \end{array} \right].\nonumber
\end{equation}
{\bf Remarks}
\begin{itemize}
	\item The matrix $A$ serves as the generator matrix of the $(r+1,r)$ MDS local code used in the second encoding step. This step provides the locality property of the code.
	\item The generator matrix $G$ is not in systematic form: no $k$ subsets of its columns form the identity matrix.
However, there is an easy way to do so, by preserving the locality and distance properties: pick $k$ linearly independent columns of $G$, say $G_k$, and use as a new code generator matrix the matrix $G_{\text{sys}} = G_k^{-1}G$.
\item {\color{black}Although Construction \ref{cnstr1} relies on using RS symbols over a finite field of order $p^{k+1}$, this requirement is not strict: we can use RS-coded symbols over a finite field of size $p$, and then group them to obtain RS symbols over $p^{k+1}$.
The details of this point are clarified in Section V.}
\end{itemize} 

%

\begin{theo}
The code generated by $G$ has locality $r$ and optimal minimum distance $d=n-k-\left\lceil \frac{k}{r}\right\rceil+2$.
\label{ththth}
\end{theo}

The proof of the above theorem is done in two steps. 
First, in Section~\ref{sec:matroids}, we derive  a new result that expresses the minimum distance of a linear code using the matroid represented by its generator matrix. This result will imply that it is sufficient to check the invertibility of a certain subset of $k\times k$ submatrices of $G$.
Then, in Section~\ref{sec:optimality}, we show that indeed each submatrix in this set is invertible; we do so by using properties of the determinant function and polynomials over finite fields.

\section{Matroids and Locally Repairable Codes}
\label{sec:matroids}

In this section, we derive a fundamental connection between a linear code and the matroid represented by the code generator matrix. 
More precisely, the minimum distance of the code will be expressed in  terms of the matroid's circuits, which we define in the following. 
Sufficient conditions for optimal locally repairable codes are derived.
We start with a brief overview of matroid theory and some basic definitions that we use throughout the paper.
We would like to emphasize that the result that connects the minimum distance of the code and the matroid represented by its generator matrix applies to any linear code (not necessarily an LRC).
We would like to mention that a similar framework was introduced in \cite{gopalan2011locality}, based on dimensionality properties of the sub-matrices of the generator matrix.

\subsection{Overview of Matroid Theory}
\label{subsec:matroid_intro}
A matroid $\mathcal{M}=\mathcal{M}([n],\rank(\cdot))$ is defined by the set of integers $[n]=\{1,...,n\}$ and an integer valued function  $\rank(\cdot)$, that is defined on all subsets of $[n]$, and  satisfies the properties:
\begin{itemize}
\item $\rank(\mathcal{A})\ge 0$, for any $\mathcal{A}\subseteq [n]$.
\item $\rank(\mathcal{A})\leq |\mathcal{A}|$, for any $\mathcal{A}\subseteq [n]$.
\item $\rank(\mathcal{A})\leq \rank(\mathcal{B})$, for any sets $\mathcal{A}\subseteq \mathcal{B} \subseteq [n]$. 
\item $\rank(\mathcal{A}\cup \mathcal{B})+\rank(\mathcal{A}\cap \mathcal{B})\leq \rank(\mathcal{A})+\rank(\mathcal{B})$, for any sets $\mathcal{A}\subseteq \mathcal{B} \subseteq [n]$. 
\end{itemize}
A set $\mathcal{A}$ is called \emph{independent} if $\rank(\mathcal{A})=|\mathcal{A}|$;
otherwise $\mathcal{A}$ is called dependent. 
A set is referred as a {\it circuit} 
if it is dependent {\it and all} of its proper subsets are independent.
This means that if a set $\mathcal{C}$ is a circuit, then $\rank(\mathcal{C}) = |\mathcal{C}|-1$.

\noindent{\bf Example:} Let $G$ be a $k\times n$ (e.g. a code generator) matrix. 
Define the matroid $\mathcal{M}([n],\rank(\cdot))$, where the rank of a set $\mathcal{A}\subseteq [n]$ is 
$\rank(\mathcal{A})=\rank(G_\mathcal{A}),$
$G_\mathcal{A}$ is the sub-matrix of $G$ with columns indexed by $\mathcal{A}$, and rank operates on a set of columns in the well-known linear-algebraic way.
In this case, the matroid $\mathcal{M}$ is said to be \emph{represented} by $G$. 

\subsection{Connections to Code Distance}
Recall that any linear code can be defined by its generator matrix or by its parity-check matrix. 
These two matrices represent two matroids that have  many interesting connections to the linear code, e.g., one can easily observe that the minimum distance of the code equals to the size of the smallest circuit in the matroid represented by the parity-check matrix. In what follows,	 we derive a new fundamental connection of this kind. More specifically, we express the minimum distance of the code using some parameters of the matroid represented by the code generator matrix. 
To the best of our knowledge, this result is new and we believe it is of independent interest, even outside the scope of LRCs.
We proceed with our technical derivations.

A collection of sets $\mathcal{C}_1,\; \mathcal{C}_2,\ldots$ is said to have a non trivial union if every set is \emph{not} contained in the union of the others, that is
 \begin{equation}
\mathcal{C}_i\nsubseteq \cup_{j\neq i}\mathcal{C}_j, \; \text{for any }i.\nonumber
 \end{equation}
Using the above definitions we state a simple lemma that will be fundamental in our derivations.
\begin{lem}
Let $\mathcal{C}_1,...,\mathcal{C}_m$ be $m$ circuits in $\mathcal{M}$. 
If the circuits have a non trivial union, then 
$$\rank\left(\bigcup_{i=1}^m\mathcal{C}_i\right)\leq \left|\bigcup_{i=1}^m\mathcal{C}_i\right|- m.$$
\label{lemma:1}
\end{lem}

 \begin{IEEEproof}
We apply induction on $m$. For $m=1$, since $\mathcal{C}_1$ is a circuit $\rank(\mathcal{C}_1)=|\mathcal{C}_1|-1.$ 
Let $m>1$ and denote by $\mathcal{C}=\bigcup_{i=1}^{m-1}\mathcal{C}_i$. By the property of the $\rank$ function
 \begin{align*}
\rank(\mathcal{C}\cup \mathcal{C}_m)\leq \rank(\mathcal{C})+\rank(\mathcal{C}_m)-\rank(\mathcal{C}\cap \mathcal{C}_m).
 \end{align*}
Since the union of the circuits is non trivial, $\mathcal{C}\cap \mathcal{C}_m$ is a proper subset of $\mathcal{C}_m$ and therefore is independent.
Then by the induction assumption 
 \begin{align*}
&\rank(\mathcal{C})+\rank(\mathcal{C}_m)-\rank(\mathcal{C}\cap \mathcal{C}_m)\leq|\mathcal{C}|-(m-1)+|\mathcal{C}_m|-1-|\mathcal{C}\cap \mathcal{C}_m|=|\mathcal{C}\cup \mathcal{C}_m|-m.
 \end{align*}
\end{IEEEproof}

%

In what follows, we consider $\cM$  to be the matroid that is represented by a code generator matrix $G$ of size $k\times n$. 
We will define a new parameter $\mu$ relevant to the matroid $\cM$, which will be used later in calculating the minimum distance of the code generated by $G$.
We would like to note that $\mu$ can be defined also for non-representable matroids as well.
We proceed with its definition and properties.
\begin{defin}
Denote by $\mu$ the minimum positive integer such that the size of every non trivial union of $\mu$ circuits in $\mathcal{M}$ is at least $k+\mu.$
\end{defin}
 The following lemma provides some properties of $\mu$.
\begin{lem}
The parameter $\mu$ is well defined and it is at most $n+1$.
\label{best lemma}
\end{lem} 

\begin{IEEEproof}
Since there is \emph{no} non trivial union of $n+1$ circuits, the statement: 
\emph{``any non trivial union of $n+1$ circuits is of size at least $k+(n+1)$"} is satisfied trivially, hence $\mu$ is the minimum of a  non-empty set.

\end{IEEEproof}

The next proposition is the main result of this section. 
It characterizes the properties of locality and minimum distance of a linear code, in terms of the circuits of the matroid represented by the code generator matrix.

\begin{prop}
Let $G$, $\cM$ and $\mu$ defined as above. Then,
\begin{enumerate}
	\item the code has locality $r$ iff each $i=1,...,n$ is contained in a circuit of size at most $r+1$,
	\item the minimum distance of the code is equal to 
$d=n-k-\mu+2.$    
\end{enumerate}
\label{good thm}
\end{prop}

\begin{IEEEproof}
\begin{enumerate}
	\item This follows trivially from the definition of a circuit.
	\item If $\mu=1$, then by Definition 1, the size of any circuit is of size at least $k+1$. Hence, any $k$ columns of $G$ are linearly independent, and $G$ is a generator matrix of  an MDS code, namely $d=n-k+1$.
If $\mu\geq 2$, then by the minimality of $\mu$ there exist $\mu-1\geq 1$ circuits $\mathcal{C}_1,...,\mathcal{C}_{\mu-1}$ whose union is non trivial, and is of size at most $k-1+\mu-1=k+\mu-2$. Let $\cC=\bigcup_{i=1}^{\mu-1}\mathcal{C}_i$.
Then, by Lemma \ref{lemma:1} 
\begin{align}
\rank(\cC)\leq |\cC|-(\mu-1)\leq k-1.
\label{xnxn}
\end{align}

Note that 
\begin{align*}
k=\rank([n])&=\rank(\cC \cup [n]\backslash \cC)+\rank(\cC \cap [n]\backslash \cC)\\
&\leq \rank(\cC)+\rank([n]\backslash \cC)\\
& \leq \rank(\cC)+|[n]\backslash \cC|,
\end{align*}
where the two inequalities follow from the properties of the $\rank$ function. Hence the size of $[n]\backslash \cC$ is at least $k-\rank(\cC).$ 
Let $\cB\subseteq [n]\backslash \cC$ be an arbitrary subset of size $k-1-\rank(\cC)$,
then
\begin{align*}
|\cB\cup \cC|&=|\cB|+| \cC|\\
&=k-1-\rank(\cC)+| \cC|\\
&\geq k-1-\rank(\cC)+\rank(\cC)+(\mu-1)\\
&= k+\mu-2,
\end{align*}
where the inequality follows from \eqref{xnxn}. Furthermore, the rank of the union of the two sets satisfies
\begin{align*}
\rank(\cB\cup \cC)&\leq \rank(\cB)+\rank(\cC)-\rank(\cB\cap \cC)\\
&\leq |\cB|+\rank(\cC)\\
&=k-1-\rank(\cC)+\rank(\cC)\\
&=k-1.
\end{align*}
Now, let $x$ be a nonzero vector of length $k$ which is orthogonal to the columns of $G$ with indices in $\cB\cup\cC$. Clearly such vector $x$ exists since the rank of $\cB\cup\cC$ is at most $k-1$. 
Then, by this choice of $x$, we get that $x\cdot G$ is a nonzero codeword of weight at most $n-(k+\mu-2)$.
Therefore, the minimum distance of the code satisfies
\begin{align*}
d\leq n-(k-\mu+2).
\end{align*}
We will now obtain a lower bound on $d$.
Let $\mathcal{T}$ be the set of zero coordinates of some nonzero codeword of the code generated by $G$ and let $\mathcal{S}\subseteq \mathcal{T}$ be a maximal independent set in $\mathcal{T}$. Clearly the size of $\mathcal{S}$ is at most $k-1$, since $\mathcal{T}$ is the set of zero coordinates of a \emph{nonzero} codeword. 
Let $\mathcal{T}\backslash \mathcal{S}=\{t_1,...,t_l\}$. 
We claim that $l\leq \mu-1$. 
Assume otherwise, then for each $i=1,...,l$ the set $t_i\cup \mathcal{S}$ is a circuit that contains $t_i$, hence the set $\mathcal{S}\cup t_1 \cup\ldots\cup t_\mu$ contains $\mu$ distinct circuits $t_i\cup \mathcal{S}$ whose union is non trivial. 
From the definition of $\mu$ we conclude that 
 \begin{align*}
k-1\geq |\mathcal{S}|
=|\mathcal{S}\cup t_1 \cup\ldots\cup t_\mu|-\mu\geq k+\mu-\mu=k,
 \end{align*}
and we get a contradiction. Therefore $l\leq \mu-1$, hence 
 the weight of the codeword is  
\begin{align*}
n-|\mathcal{T}|=n-(|\mathcal{S}|+|\mathcal{T}\backslash \mathcal{S}|)\geq n-(k-1+\mu-1).
\end{align*}
This implies that the minimum distance is at least $d\geq n-k-\mu+2$.
The result follows by combining the upper and lower bounds on $d$. 
\end{enumerate}
\end{IEEEproof}
Observe that the second part of the proposition, which provides a characterization of the minimum distance, does not use any assumptions on locality, and thus it applies to \emph{any} linear code.
This is the main point that differentiates our use of matroids, from the related framework of \cite{gopalan2011locality}.

%
From Proposition \ref{good thm}, we get the following theorem which characterizes all optimal linear LRCs.
\begin{theo}
The code generated by $G$ has locality $r$ and optimal minimum distance $d=n-(k+\lceil \frac{k}{r}\rceil)+2$, if and only if,
\begin{enumerate}
	\item any $i=1,...,n$ is contained in a circuit of size at most $r+1$, \text{ and }
	\item the size of any nontrivial union of $\lceil \frac{k}{r}\rceil$ circuits in  $\mathcal{M}$ is at least $k+\lceil \frac{k}{r}\rceil.$  	
\end{enumerate}
\label{best cor}
\end{theo}

 \begin{IEEEproof}
 By the above two conditions and Proposition \ref{good thm} we conclude that the code generated by $G$ has locality $r$ and minimum distance at least $n-k-\lceil \frac{k}{r}\rceil+2$. However, it was shown in \cite{gopalan2011locality,papailiopoulos2012simple} that a code with locality $r$ has minimum distance at most $n-k-\lceil \frac{k}{r}\rceil+2$. 
 \end{IEEEproof}
The previous theorem provided necessary and sufficient conditions for an optimal linear LRCs. 
The following corollary gives a simple necessary conditions for optimal linear LRC. 
This corollary will be used in the next section in order to prove the optimality of the code construction.

In what follows, we call a circuit nontrivial if its size is at most $k$, and trivial otherwise.
\begin{cor}
\label{cor1}
Let $G$ and $\mathcal{M}$ as before, then the code has locality $r$, and optimal minimum distance $d=n-k-\lceil \frac{k}{r}\rceil+2$, if
all \emph{nontrivial circuits} are of size $r+1$, and they form a partition of $[n].$
\label{good cor}  
\end{cor}

 \begin{IEEEproof}
 The locality follows since by assumption each $i=1,...,n$ is contained in a circuit of size at most $r+1$. Let $\cC_1,...,\cC_{\lceil \frac{k}{r}\rceil}$ be a collection of $\lceil \frac{k}{r}\rceil$ circuits of $\mathcal{M}$ whose union is non trivial. If one of the circuits is trivial, say $\cC_1$, then we get
  \begin{align*}
 \left|\bigcup_{i=1}^{\lceil \frac{k}{r}\rceil}\mathcal{C}_i\right|=|\mathcal{C}_1|+\sum_{i=2}^{\lceil \frac{k}{r}\rceil}\left|\mathcal{C}_i\backslash \bigcup_{j=1}^{i-1}\mathcal{C}_j\right|
 \geq k+1 +\sum_{i=2}^{\lceil \frac{k}{r}\rceil}1=k+\left\lceil \frac{k}{r}\right\rceil,
  \end{align*} 
where the inequality follows since the union of the circuits is nontrivial. 
If all the $\cC_i$'s are non trivial circuits, then it is clear that 
$$\left|\bigcup_{i=1}^{\lceil \frac{k}{r}\rceil}\mathcal{C}_i\right|=\sum_{i=1}^{\lceil \frac{k}{r}\rceil}|\mathcal{C}_i|=\left\lceil \frac{k}{r}\right\rceil(r+1)\geq k+\left\lceil \frac{k}{r}\right\rceil,$$
and the result follows from Theorem \ref{best cor}.
 \end{IEEEproof}

\section{Optimality of  the Code Construction}
\label{sec:optimality}
In this section we prove Theorem \ref{ththth}. This will be done by using Corollary \ref{good cor}, namely, showing that the nontrivial circuits of $\cM$ are of size $r+1$, and they form a partition of $[n]$. Each circuit of size $r+1$ in $\cM$ corresponds to a different repair group of symbols of size $r+1$. 
We proceed with our technical derivations. 

For $i=1,\ldots,m/r$, let $V_i$ be the Vandermonde matrix of size $k\times r$ defined by the elements 
$\alpha_{1+r(i-1)},...,\alpha_{ir}\in \mathbb{F}_p$,  
$$V_i=\left( \begin{array}{cccc}
\overline{\alpha}_{r(i-1)+1},&\overline{\alpha}_{r(i-1)+2},  & \ldots & ,\overline{\alpha}_{ir} \\
\end{array} \right),$$
where $\overline{\alpha}_i=(1,\alpha_i,\ldots,\alpha_i^{k-1})^t.$
Then, we can rewrite the generator matrix $G$ in \eqref{eq:ththth} as
\begin{equation*}
G=(V_1\cdot A,V_2\cdot A,\ldots,V_{m/r}\cdot A),
\end{equation*}
where $A$ is the $r\times (r+1)$ generator matrix of the $(r+1,r)$-MDS local code.
{\color{black}
\begin{lem}
The code generated by $G$ has locality $r$, and for any $i=1,\ldots,\frac{n}{r+1}$, the set $\mathcal{C}_i=[1+(i-1)(r+1), \ldots, i(r+1)]$ forms a circuit of size $r+1$ in the matroid $\cM$.
\label{eded}
\end{lem}
\begin{IEEEproof}
Since $A$ is a generator matrix of an $(r+1,r)$-MDS code, on the input of $r$ symbols it generates $r+1$ symbols such that each symbol can be repaired
by accessing the $r$ remaining symbols that come from the same $(r+1,r)$-MDS local code.
Moreover, since any $r$ columns of $A$ are linearly independent and the $r+1$ columns of $A$ are dependent, we get that each $\cC_i$ is a circuit in $\cM$.
\end{IEEEproof}
It is clear that the circuits $\cC_i$ form a partition of $[n]$, hence in order to establish the optimality of the distance, we will show that these are the \emph{only} nontrivial circuits of $\cM$. 
This will be proved in Lemma \ref{main lemma}, but in order to do that, we need to extend the definition of the permanent function, and to establish an important property of the matrix $A$.

The permanent function is defined for square matrices, however the definition can be naturally extended to non-square matrices as follows. 
\begin{defin}
Let $B=(b_{i,j})$ be an $r\times t$ matrix and $t\leq r$, then
\begin{equation}
\perm(B)=\sum_{\substack{(v_1,\ldots,v_t)\\1\leq v_i\neq v_j\leq r}}\prod_{i=1}^tb_{v_i,i}.
\label{eq:987}
\end{equation}
\end{defin}

Intuitively, the permanent is the sum of all possible products of elements in $B$, such that exactly one entry is picked from each column, and no two elements are picked from the same row. 
In the sequel we will calculate the permanent of submatrices of $A$ defined in Construction \ref{cnstr1}, which is a matrix with entries $0,1$ and $\omega$. 
For the calculation of the permanent of any submatrix of $A$, we will view the matrix as a matrix over $\mathbb{Z}[\omega]$, the ring of polynomials in the variable $\omega$ over the integers. In other words, each entry of the matrix $A$ is viewed as a polynomial in $\mathbb{Z}[\omega]$.
This will imply that the value of the permanent function is a polynomial in $\mathbb{Z}[\omega]$.   
In order to make our point clear, consider the following matrix 
\[B= \left( \begin{smallmatrix}
1& 0   \\
\omega & 1   \\
0& \omega   \\
 \end{smallmatrix} \right).\]
To calculate the permanent of $B$, we consider $\omega$ as a variable, and then we have
\begin{align*}
\perm(B)&=\sum_{\substack{1\leq v_1\neq v_2\leq 3}}\prod_{i=1}^2b_{v_i,i}\\
&= b_{1,1}(b_{2,2}+b_{3,2})+b_{2,1}(b_{1,2}+b_{3,2})+b_{3,1}(b_{1,2}+b_{2,2})\\
&=1\cdot(1+\omega)+\omega\cdot(0+\omega)+0\cdot(0+1)\\
&=\omega^2+\omega+1.
\end{align*}
We proceed with an important property of the matrix $A$.
\begin{lem}
Let $B$ be an $r\times t$ sub-matrix of $A$ for $r\geq t$, then the permanent of $B$ is a monic polynomial in $\omega$ of degree at most $t$. 
\label{dimitris dimitris}
\end{lem}

\begin{IEEEproof}
By the structure of the matrix $A$ it is evident that $B$ is a block diagonal matrix with blocks $B_1,\ldots,B_m$ for some $m$, and each matrix $B_i$ is composed of consecutive columns of $A$.
%
Hence the permanent of $B$ is the product of the permanent of its blocks, and the permanent of $B$ is a monic polynomial if the permanent of each block matrix $B_i$ is a monic polynomial. If the matrix $B_i$ contains the first column of the matrix $A$, namely $B_i$ is composed of the first $l$ columns of $A$ for some integer $l$, then $\perm(B_i)=1$, which is clearly a monic polynomial. If $B_i$ does not contain the first column of $A$, then $B_i$ is composed of the columns of $A$ with indices in the set $[l_1,\ldots,l_2]$, and $2\leq l_1\leq  l_2$. In this case one can verify that $\perm(B_i)$ is a monic polynomial of degree $l_2-l_1+1$. We conclude that the permanent of each of the block matrices $B_i$ is a monic polynomial, and hence also the permanent of $B$.
For the second part note that each of the summands in \eqref{eq:987} is a product of exactly $t$ elements of $B$. In addition, each element equals to  $\omega, 1$ or $0$, hence the degree of each term is at most $t$, and the result follows.
\end{IEEEproof}
In order to clarify the previous lemma consider the matrix $B$ which is composed of columns $1, 2, 4$ and $5$ of $A$ of size $4\times 5$, then
\[B= \left( \begin{smallmatrix}
1& \omega  & 0& 0 \\
0& 1  & 0& 0 \\
0& 0  & \omega& 0 \\
0& 0  & 1& \omega \\
 \end{smallmatrix} \right).\]
 Let 
$B_1= \left( \begin{smallmatrix}
1& \omega   \\
0& 1  
\end{smallmatrix} \right)\text{ and } B_2= \left( \begin{smallmatrix}
\omega & 0   \\
1& \omega  
\end{smallmatrix} \right).$
Then, 
$\perm(B)=\perm(B_1)\cdot\perm(B_2)=1\cdot \omega^2=\omega^2,$ which is a monic polynomial.

}

Now we are ready to prove the lemma on the nontrivial circuits of $\cM$.
\begin{lem}
\label{main lemma}
$\cC_1,\ldots,\cC_{\frac{n}{r+1}}$ are the only nontrivial circuits of $\cM$.
\end{lem}

\begin{IEEEproof}
{\color{black}Let $\mathbb{S}$ be all the $k$-subsets of $[n]$ that do \emph{not} contain any circuit $\mathcal{C}_i$, namely 
\begin{equation}
\mathbb{S}=\{\mathcal{S}\subseteq [n]: |\mathcal{S}|=k \text{ and } \mathcal{C}_i \nsubseteq \mathcal{S} \text{ for any } i=1,\ldots,\frac{n}{r+1}  \}.
\label{rcrc}
\end{equation}

For $\mathcal{S}\in \mathbb{S}$, denote by $G_\mathcal{S}$ the square sub-matrix of $G$ restricted to columns with indices in $\mathcal{S}$. 
It is clear that proving the claim is equivalent to proving that any submatrix $G_\cS$ for $\cS\in \mathbb{S}$ is invertible. This will be done by showing that the determinant of $G_\mathcal{S}$ is a \emph{nonzero} polynomial in $\omega$, of degree at most $k$, and with coefficients in $\mathbb{F}_p$. 
Since $\omega$ is a primitive element of the field $\mathbb{F}_{p^{k+1}}$, the degree of its minimum polynomial in $\mathbb{F}_p[x]$ is \emph{exactly} $k+1$. 
This will imply that the determinant of $G_\cS$ is a nonzero element of $\mathbb{F}_{p^{k+1}}$, namely $G_\mathcal{S}$ is invertible, and the result will follow.
}


We will first present an example for the case of $r=3,\; k=6$, and then proceed with the general statement.
Let $G$ be the generator matrix of an $(n,6,3)$ LRC of Construction \ref{cnstr1}, i.e.
$$G=(V_1\cdot A,\ldots,V_{n/4}\cdot A),$$
and $A$ is an $3\times 4$ generator matrix of the $(4,3)$-MDS local code.
The circuits of size $4$ of $\cM$ that partition the set $[n]$ are $\cC_1=[1,\ldots,4],\cC_2=[5,\ldots,8],\ldots,\cC_{n/4}=[n-3,\ldots,n]$. Consider the set $\cS=\{1,2,3,5,7,10\}\in \mathbb{S}$, and note that $\cC_i\nsubseteq \cS$ for any $i$.
The matrix $G_\cS$ contains the first three columns of $V_1\cdot A$, the first and third column of $V_2\cdot A$, and the second column of $V_3\cdot A$.
For $i=1,2,3$ define $A_i$ a sub-matrix of $A$,
$$A_1=\left( \begin{smallmatrix}
1& \omega & 0   \\
0& 1 & \omega\\
0& 0 & 1
\end{smallmatrix} \right),
A_2=\left( \begin{smallmatrix}
1&  0   \\
0& \omega\\
0& 1
\end{smallmatrix} \right),\text{ and }
A_3=\left( \begin{smallmatrix}
\omega   \\
1 \\
0 
\end{smallmatrix} \right).
$$
Then $G_\mathcal{S}$ can be written as 
\begin{align}
G_\mathcal{S}=(V_{1}A_{1},V_2A_2,V_{3}A_{3})&=(V_1,V_2,V_{3})\cdot D(A_{1},A_2,A_{3})\nonumber\\
&=(\overline{\alpha}_1, \omega \overline{\alpha}_1+\overline{\alpha}_2, \omega \overline{\alpha}_2+\overline{\alpha}_3, \overline{\alpha}_4, \omega \overline{\alpha}_5+\overline{\alpha}_6, \omega \overline{\alpha}_7+\overline{\alpha}_8). \label{tvtv}
\end{align}
Where $D(A_1,\ldots,A_t)$ is a block diagonal matrix with the matrices $A_i$ along its diagonal.
Consider the determinant of $G_{\mathcal{S}}$, and recall that the determinant function is linear in the columns of a matrix: if $u$ and $v$ are column vectors, $B$ is some matrix, and $\alpha,\beta$ are scalars, then 
$$\det([\alpha \cdot v+\beta \cdot u,\;B])=\alpha \cdot \det([v,\;B])+ \beta \cdot \det([u,\;B]).$$
Using linearity, the determinant of $G_\mathcal{S}$ in (\ref{tvtv}) can be expanded into a linear combination of powers of $\omega$ multiplied by determinants of Vandermonde matrices, where each Vandermonde matrix is defined by elements from the field $\mathbb{F}_p$. In other words, the determinant of
$G_\mathcal{S}$ is a polynomial in $\omega$ over $\mathbb{F}_p$. Note that in \eqref{tvtv}, the second, third, fifth, and sixth columns  of $G_\mathcal{S}$ are a linear combination of two  
columns, hence in the expansion of the determinant we will get $2^4$ distinct summands, depending on which columns we used to  expand the determinant. Many of the summands will be equal to zero, e.g., we get the following term  in the expansion of $\det(G_\mathcal{S})$ 
\begin{equation}
\omega^4\cdot\det(\overline{\alpha}_1,\;\overline{\alpha}_1,\;\overline{\alpha}_2,\;\overline{\alpha}_4,\;\overline{\alpha}_5,\;\overline{\alpha}_7),
\label{finally}
\end{equation}
 which equals to zero since the column $\overline{\alpha}_1$ appears twice in \eqref{finally}.
Clearly, the only Vandermonde matrices in the expansion of $\det(G_\cS)$ that contribute a nonzero term, are those who have distinct columns. 
One can check that  
\begin{align}
\det(G_\cS)=& \omega^2\cdot\det(\overline{\alpha}_1,\;\overline{\alpha}_2,\;\overline{\alpha}_3,\;\overline{\alpha}_4,\;\overline{\alpha}_5,\;\overline{\alpha}_7)+\nonumber\\
&\omega\cdot(\det(\overline{\alpha}_1,\;\overline{\alpha}_2,\;\overline{\alpha}_3,\;\overline{\alpha}_4,\;\overline{\alpha}_5,\;\overline{\alpha}_8)+
\det(\overline{\alpha}_1,\;\overline{\alpha}_2,\;\overline{\alpha}_3,\;\overline{\alpha}_4,\;\overline{\alpha}_6,\;\overline{\alpha}_7))+\label{ujuj1}\\
&1\cdot\det(\overline{\alpha}_1,\;\overline{\alpha}_2,\;\overline{\alpha}_3,\;\overline{\alpha}_4,\;\overline{\alpha}_6,\;\overline{\alpha}_8)\nonumber.
\end{align}
It is easy to see that the number of nonzero summands in the coefficient of $\omega^i$ for $i=0,1,2$ equals to the coefficient of $\omega^i$ in the permanent of $D(A_1,A_2,A_3)$. For example, if 
$$\perm(D(A_1,A_2,A_3))=\sum_{i=0}^2\beta_i\omega^i,$$
and $\beta_i\in \mathbb{Z}$, then since there are two nonzero summands in the coefficients of $\omega$ in \eqref{ujuj1} we conclude that $\beta_1=2$. Similarly, $\beta_0=\beta_2=1$. Since the $A_i$'s are submatrices of $A$, then by Lemma \ref{dimitris dimitris}  $\perm(D(A_1,A_2,A_3))$ is a monic polynomial, hence the leading coefficient of $\det(G_\cS)$ is the sum of exactly \emph{one} nonzero term (the term $\det(\overline{\alpha}_1,\;\overline{\alpha}_2,\;\overline{\alpha}_3,\;\overline{\alpha}_4,\;\overline{\alpha}_5,\;\overline{\alpha}_7)$). Hence the leading coefficient of $\det(G_\cS)$ is nonzero, since the sum of one nonzero number is nonzero. We conclude that for the example above, $\det(G_\cS)$ is a non zero polynomial in $\omega$ with coefficients in $\mathbb{F}_p$. 


In the general case $G_\mathcal{S}$ can be written as   
$$G_\mathcal{S}=(V_{i_1}A_{1},\ldots,V_{i_t}A_{t})=(V_{i_1},\ldots,V_{i_t})\cdot D(A_1,\ldots,A_t)$$ 
for some $1\leq t\leq m/r$, and $D=D(A_1,\ldots,A_t)$ is a block diagonal matrix with the matrices $A_i$ along its diagonal.
Here, each $A_i$ is again a sub-matrix of $A$.
The coefficient of $\omega^i$ in $\perm(D)$ equals to the number of nonzero terms of degree $i$ in the expansion of $\det(G_\mathcal{S})$. By Lemma \ref{dimitris dimitris}, each of the polynomials $\perm(A_i)$ is monic.
This also implies that $\perm(D)$ is a monic polynomial, since the permanent of $D$ equals to the product of the permanent of its blocks.
Hence, there is only one non zero term in the expansion of $\det(G_\mathcal{S})$ with the largest degree of $\omega$. Namely, $\det(G_\mathcal{S})$ is a nonzero polynomial in $\omega$ over $\mathbb{F}_p$. 
Moreover, $D$ has $k$ columns, therefore the determinant is a non zero polynomial of degree at most $k$. 
For the final step, since the minimum degree of a non zero polynomial over $\mathbb{F}_p$ that annihilates $\omega$ is $k+1$, we conclude that $\det(G_\mathcal{S})$ is a non zero element in $\mathbb{F}_{p^{k+1}}$, and therefore $G_\mathcal{S}$ is invertible.

\end{IEEEproof}

%

By combining  Lemma \ref{eded} and Lemma \ref{main lemma} the nontrivial circuits of $\cM$ are of size $r+1$, and they form a partition of $[n]$. Then by Corollary \ref{good cor} we conclude that $G$ generates an optimal $(n,k,r)$ LRC, and Theorem \ref{ththth} is established.

\section{Efficient Encoding, and Almost-MDS Distance}
\label{fast_encoding}
\subsection{Encoding top of existing Reed-Solomon stripes}
Construction \ref{cnstr1} has a very important property: the coded symbols can be generated using already coded Reed-Solomon symbols.
This design flexibility of the presented codes comes in sharp contrast to other schemes, which require to decode the entire already stored information; a process that can be cost inefficient, when it comes to large scale storage applications.
{\color{black}In our case, when RS-symbols are already stored in a system, the only coding overhead is in re-encoding groups of $r$ RS-coded symbols, in $r+1$ coded symbols, which can be done with complexity $O(n)$, since $A$ has at most $2$ elements per row and column.}


Recall that the first step encoding of the construction generated RS encoded symbols over $\mathbb{F}_{p^{k+1}}$, where the evaluation points are taken from the subfield $\mathbb{F}_p.$ In other words, we evaluate a polynomial of $\mathbb{F}_{p^{k+1}}[x]$ at the points of the field $\mathbb{F}_p$. The following lemma shows that grouping together $k+1$ RS encoded symbols over $\mathbb{F}_p$ can be viewed as a single encoded RS symbol over $\mathbb{F}_p$. Hence, already coded RS symbols over $\mathbb{F}_p$ can be used in the first encoding step, by a simple grouping, and the entire construction reduces to performing only the second step of the local encoding.

\begin{lem}
Each coded symbol of an $(m,k)$-RS code over $\mathbb{F}_{p^{k+1}}$ evaluated at points of the field $\mathbb{F}_p$ has an equivalent representation as a vector of $k+1$ coded symbols of an $(m,k)$-RS code over $\mathbb{F}_{p}$
\end{lem}

\begin{IEEEproof}
Let $\beta=(\beta_0,\ldots,\beta_{k-1})$ be the information symbols to be encoded, where each $\beta_i\in \mathbb{F}_{p^{k+1}}.$ Let $\omega$ be a primitive element of 
$\mathbb{F}_{p^{k+1}}$, then each symbol $\beta_i$ can be written as 
a polynomial in $\omega$ of degree at most $k$ and coefficients from $\mathbb{F}_p$, namely
$$\beta_i=\sum_{j=0}^{k}\beta_{i,j}\omega^{j}, \text{ and } \beta_{i,j}\in \mathbb{F}_p.$$
The RS symbol over $\mathbb{F}_{p^{k+1}}$ is simply the evaluation of the polynomial $f_\beta(x)$ defined as $$f_\beta(x)=\sum_{i=0}^{k-1}\beta_ix^{i},$$
at $m$ distinct elements of the field $\mathbb{F}_p$. 
However
\begin{align*}
f_\beta(x)&=\sum_{i=0}^{k-1}\beta_ix^{i}=\sum_{i=0}^{k-1}\sum_{j=0}^{k}\beta_{i,j}\omega^{j}x^{i}=\sum_{j=0}^{k}\omega^{j}\sum_{i=0}^{k-1}\beta_{i,j}x^{i}=\sum_{j=0}^{k}\omega^{j}g_j(x),
\end{align*}
where $g_j(x)$ is a polynomial of degree at most $k-1$ over $\mathbb{F}_p$.
Specifically for $\alpha\in \mathbb{F}_p$  
\begin{equation}
f_\beta(\alpha)=\sum_{j=0}^{k}\omega^{j}g_j(\alpha).
\label{tgtgtg}
\end{equation}
In other words, the evaluation of $f_\beta(x)$ at the point $\alpha\in \mathbb{F}_p$ equals to the summation of $k+1$ summands, and each summand is a product of some power of $\omega$ with an evaluation of a polynomial of degree less than $k$ over $\mathbb{F}_p$. Therefore, if  $g_j(\alpha)$ for $j=0,\ldots,k$ are $k+1$ RS encoded symbols derived by evaluating $k+1$ polynomials $g_j(x)$ over $\mathbb{F}_p$ at the point $\alpha \in \mathbb{F}_p$, then \eqref{tgtgtg} can be viewed as a one encoded RS symbol over $\mathbb{F}_{p^{k+1}}$ evaluated at $\alpha \in \mathbb{F}_p$. 
This symbol can be represented as a vector of length $k+1$ over $\mathbb{F}_p$: this is done by a simple concatenation of the $k+1$ symbols into the vector $(g_0(\alpha),\ldots,g_{k}(\alpha))$.
\end{IEEEproof}
\begin{figure}[h]
\centerline{
\includegraphics[width=0.7\columnwidth]{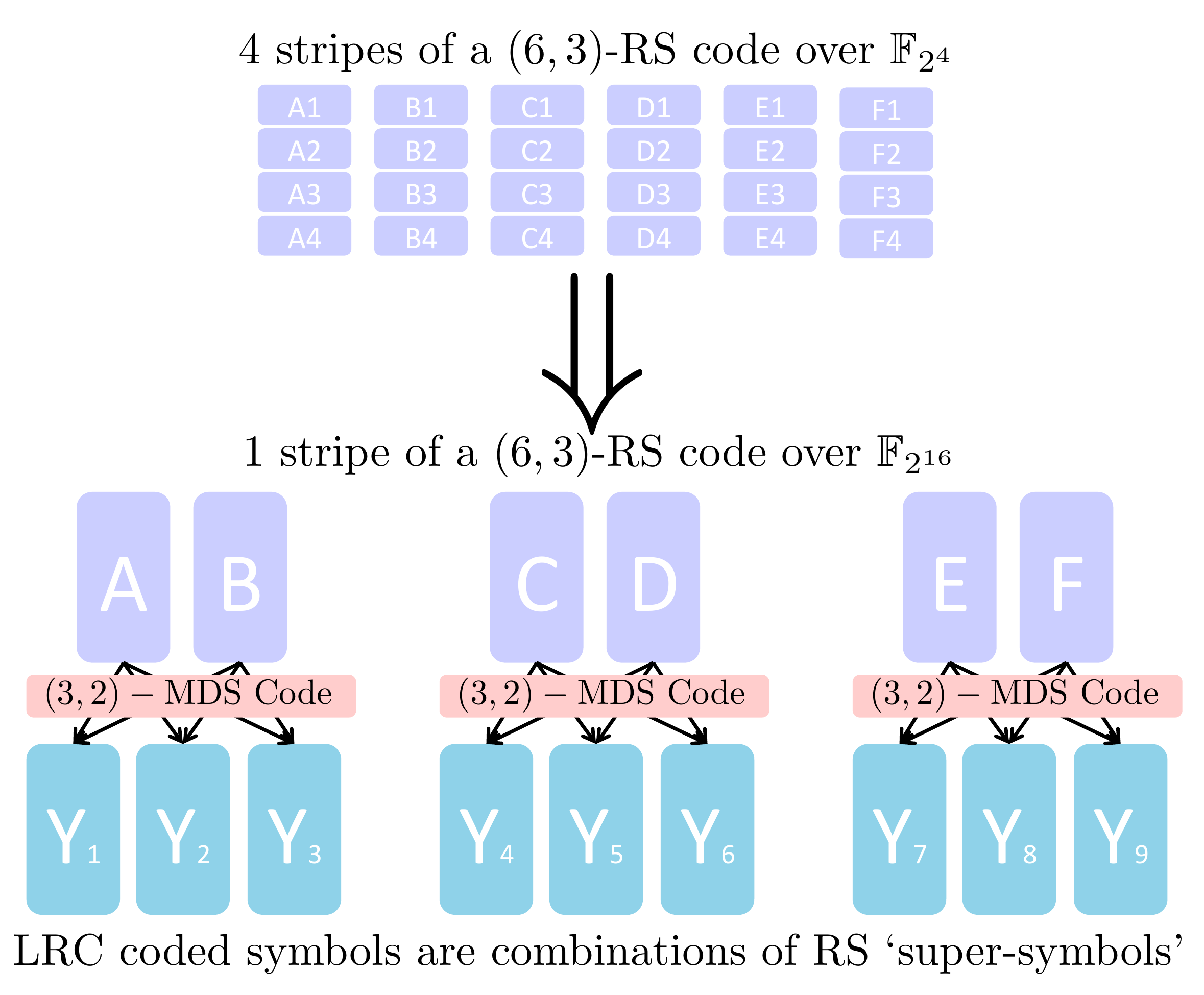}
}
\caption{Example of a $(n=9,k=3,r=2)$ LRC using Construction \ref{cnstr1}. Assume that $k+1=4$ stripes of RS-encoded data over $\mathbb{F}_{2^4}$ are stored in $6$ nodes. Each node can be viewed as storing a single RS symbol over $\mathbb{F}_{(2^4)^4}$. The second step encoding partitions the symbols into pairs, and generates from any pair, $3$ symbols over $\mathbb{F}_{2^{16}}$, to provide the locality property.}
\label{fig:super_symbols}
\end{figure}
This observation is particularly useful in the following array setting: there are $n$ nodes, and the $i$-th coded symbols of $k+1$ RS coded outputs evaluated at the same point, are stored in node $i$. Then, node $i$ can be equivalently seen as storing $1$ RS coded symbol over $\mathbb{F}_{p^{k+1}}$.
In Fig.~\ref{fig:super_symbols}, we give an example of an $(9,3,2)$ LRC. In the first step of the encoding we evaluate the polynomial at $m=\frac{nr}{r+1}=6$ distinct points of the field $\mathbb{F}_p$, hence $p\geq 6$. Assuming that we would like to operate over binary characteristic, we pick $p=2^4$ (note that also $p=2^3$ would suffice). Let $\alpha_1,\ldots,\alpha_6\in \mathbb{F}_{2^4}$ be six distinct elements, and assume that the $k+1=4$ rows in Figure \ref{fig:super_symbols} represent four different stripes of an $(6,3)-$RS code, evaluated at the points $\alpha_i$. Moreover, each column corresponds to an evaluation at a different point, e.g., the second column $(B1,B2,B3,B4)^t$ corresponds to the evaluation at the point $\alpha_2$. By the previous lemma, the symbols of the $i$-th column can be viewed as an evaluation of a polynomial over $\mathbb{F}_{(2^4)^4}=\mathbb{F}_{2^{16}}$ at the point $\alpha_i$. In other words, for the first encoding step we only need to group together an already RS coded symbols, and the only computational task is in the second step, that provides the locality property.

\subsection{Decoding beyond the minimum distance: tolerating asymptotically as many erasures as an MDS code.}
In this subsection, we show that Construction \ref{cnstr1} has decoding capabilities far beyond its minimum distance, in most cases of erasure patterns.
It is well known that a code of distance $d$, can tolerate in the worst case at most $d-1$ symbol erasures.
Observe that Construction \ref{cnstr1} has distance $d=n-k-\left\lceil\frac{k}{r}\right\rceil+2$: it can tolerate at most $n-k-\left\lceil\frac{k}{r}\right\rceil+1$ erasures, that is $\left\lceil\frac{k}{r}\right\rceil-1$, less than an $(n,k)$ MDS code.
In this subsection, we show a surprising property of Construction \ref{cnstr1}:  although it has smaller distance than an $(n,k)$ MDS code, we can still recover the information from most scenarios of $n-k$ erasures. 
In the following we make the reasonable assumption that all coded symbols can be erased with equal probability.
Under this assumption, we show the following theorem.
\begin{theo}
Let $\frac{k}{n} = R$ be fixed, and $r=\Omega(\log(n))$. 
Then, the probability that we can reconstruct the file from $k$ randomly picked symbols, goes to $1$ as $n,k\rightarrow\infty$. 
\end{theo}

\begin{IEEEproof}
We aim to calculate the probability of the following event: ``$k$ coded symbols, selected uniformly at random, can recover the file".
To do so, we can equivalently ask the following question: ``what is the probability that $k$ columns of $G$ are full-rank $k$?".
We start by enumerating all $k$-subsets of columns of $G$ that are not full-rank.
Certainly, a $k$-subset of columns that contains all $r+1$ columns of a local repair group (also referred to as a circuit in terms of the matroid language) is rank deficient.
We claim that these $k$-subsets containing ``trivial" linear dependencies, are the only ones that can lead to failure to decode.
This follows from Lemma \ref{main lemma}, where we showed that any submatrix $G_\cS$ of $G$, where $\cS$ belongs to the set defined in \eqref{rcrc}, is of full rank.

Denote by $P_{\text{dec}}$ the probability that the chosen set of symbols suffices for decoding. 
This probability equals to the probability that a uniformly chosen $k$-subset of $[n]$ does not contain any of the circuits $\mathcal{C}_1,\ldots,\mathcal{C}_{\frac{n}{r+1}}$. This probability can be easily calculated using the inclusion-exclusion principle \cite{wilson}. Let $A_i$ be the event that the chosen $k$-subset contains the circuit $\mathcal{C}_i$, then the probability is 
$$P_{\text{dec}}=1-P(\cup_iA_i)=1-\frac{\sum_{j=1}^{\lfloor \frac{k}{r+1}\rfloor}(-1)^j\binom{n/(r+1)}{j}\binom{n-j(r+1)}{k-j(r+1)}}{\binom{n}{k}}.$$
Although the above expression gives the exact value of the probability of successful decoding, we will provide a much more convenient expression derived using the union bound:
$$P_{\text{dec}}=1-P(\cup_iA_i)\geq 1-\sum_iP(A_i)= 1- \frac{n}{r+1}\frac{\binom{n-(r+1)}{k-(r+1)}}{\binom{n}{k}}\geq 1-\frac{n}{r+1}\left(\frac{k}{n}\right)^{r+1}.$$
Now, when $r\geq \frac{c\log n}{\log 1/R}$, and $c>1$ then it is not hard to see that  $P_{\text{dec}}$ goes to $1$ as $k,n\rightarrow\infty$, since 
$$\frac{n}{r+1}\left(\frac{k}{n}\right)^{r+1}\leq 
2^{\log n+r\log R}\overset{n,k\rightarrow \infty}{\longrightarrow}0.$$  
Hence, in an $(n,k,r)$ LRC of Construction \ref{cnstr1}, when $r=\Omega(\log(n))$, a fraction of $1-o(1)$ of all the $k$-subsets of coded symbols can decode the file.
\end{IEEEproof}

\section{Generalization of the Construction}
\label{Generalization}
In this section, we generalize our previous construction to one that offers local repair even if there are multiple failures coming from the same repair group. 
We accomplish that with the use of extra local parities.
We present LRCs where each symbol is contained in an MDS local code with minimum distance $\delta$. 
More formally, an $(n,k,r,\delta)$ LRC is a code such that each symbol in contained in an $(r+\delta-1,r)$ MDS local code. Therefore each symbol can be locally repaired by accessing any $r$ of the remaining $r+\delta -2$ of its local code. In other words, each local code can tolerate up to $\delta-1$ local erasures simultaneously.
For this scenario of multiple local erasures, the distance bounds of \cite{gopalan2011locality} and \cite{papailiopoulos2012locally}, were generalized in \cite{kamath2012codes}. It was shown that the minimum distance of any $(n,k,r,\delta)$ LRC 
satisfies
\begin{equation}
d\leq n-k-\left(\left(\left\lceil \frac{k}{r} \right\rceil -1\right)\left(\delta-1\right)+1\right)+2.
\label{eq:rcrc}
\end{equation}
Note that the bound in \eqref{eq:rcrc} reduces to the bound in \eqref{eq:gopalan} for $\delta=2$. 
We will now construct codes that achieve this bound.
Recall that in our previous construction, we use two encoding steps. 
For the first step, we encode the information symbols using an RS-code.
For the second step, we re-encode any $r$ RS encoded symbols into $r+1$ symbols using a specific matrix $A$ that generates an $(r+1,r)$ MDS local code. 
In the generalized construction, we modify only the second step by using  a different matrix $A$ that generates an $(r+\delta -1,r)$ MDS local code. 

Let $A$ be an $r\times (r+\delta -1)$ matrix such that its entries equal to zero or powers of the variable $\omega$, such that the permanent of any $r\times t$ submatrix of $A$ for $t\leq r$ is a \emph{monic} polynomial in $\omega$. 
We can easily construct such a matrix by using  sufficiently large powers of $\omega$. One such matrix $A$ is defined as: the $(i,j)$-th entry equals to $\omega^{(i-1)r^{j}}$. E.g., for $r=\delta=3$ we get the following matrix
\[A= \left( \begin{array}{ccccc}
\omega^0					& \omega^{0}  				& \omega^{0}& \omega^{0}& \omega^{0}\\
\omega^{ 3}& \omega^{3^2}  & \omega^{3^3}& \omega^{3^4}& \omega^{3^5}\\
\omega^{2\cdot 3}& \omega^{2\cdot 3^2}  & \omega^{2\cdot 3^3}& \omega^{2\cdot 3^4}& \omega^{2\cdot 3^5}\\
 \end{array} \right).\]
One can easily check that the permanent of any $r\times t$ submatrix of $A$ for $t\leq r$, is a monic polynomial in $\omega$.
Now, let us denote by $a$ the largest degree of $\omega$  among the entries of the matrix $A$, and  and let $\mathbb{F}_p$ be a field of size $p\geq m=\frac{nr}{r+\delta -1}$.
Assume that  $\omega$  is a primitive element of the field extension  
$\mathbb{F}_{p^{ka+1}}$. Using $m$ distinct elements $\alpha_i$ of the field $\mathbb{F}_p$, define $m/r$ Vandermonde matrices $V_i$ of order $k\times r$, as 
$$V_i=\left( \begin{array}{cccc}
\overline{\alpha}_{r(i-1)+1},&\overline{\alpha}_{r(i-1)+2},  & \ldots & ,\overline{\alpha}_{ir} \\
\end{array} \right).$$
Next, define the generator matrix of the code to be 
$$G=(V_1 A,\ldots,V_\frac{m}{r}A),$$
%
and let $\cM$ be the matroid represented by it. We claim that $G$ generates an optimal $(n,k,r,\delta)$ LRC. 
First we want to show that indeed $A$ generates an $(r+\delta -1,r)$ MDS code. In other words, any $r\times r$ submatrix of $A$ is invertible. 
Consider one such $r\times r$ submatrix $B$, and recall the difference between the definition of the determinant and the permanent of a square matrix. Notice that since $\perm(B)$ is monic, its
leading term appears also as a leading term in the determinant of $B$ (maybe with a minus sign). Moreover, since each entry of $A$ is $\omega^i$ for some $i\geq 0$ or zero, the determinant of $B$ is a polynomial in $\omega$ of degree at most $ra< ka+1$ over $\mathbb{F}_p$. Since the leading coefficient in the calculation of the determinant of $B$ is $1$ or $-1$, we conclude that it is a nonzero polynomial of degree less than $ka+1$, hence it can not annihilate $\omega$, namely, the determinant is nonzero, and the locality property of the code follows.  All that is left to be shown is the optimality of the minimum distance. For that we will need the following lemma, but first recall that a nontrivial circuit in $\cM$ is a circuit of size at most $k$.
\begin{lem}
In the matroid $\cM$, there are no non-trivial circuits, except the circuits within each local code.
Equivalently,  if $\mathcal{C}\subseteq [n]$ is a non-trivial circuit, then there exists an $i=1,\ldots,n/(r+\delta -1)$ such that $$\mathcal{C}\subseteq \left[(r+\delta-1)(i-1)+1,\ldots,(r+\delta-1)i\right].$$
\label{rfrf}
\end{lem}

\begin{IEEEproof}
By the construction of the matrix $G$, for any $i=1,\ldots,n/(r+\delta -1)$, the local code restricted to coordinates with indices in $\mathcal{C}_i=\left[1+(i-1)(r+\delta -1),\ldots,i(r+\delta -1)\right]$ is an $(r+\delta-1,r)$ MDS code. 
Therefore any subset of indices of $\mathcal{C}$ of size $r+1$ that is contained in some $\mathcal{C}_i$ forms a non-trivial circuit. 
We will now show that these are the only non trivial circuits in $\cM$.
The proof follows along the same lines as the proof of Lemma~\ref{main lemma}. 
Let $\mathbb{S}$ be all the $k$-subsets of $[n]$ that contain at most $r$ indices from each set $\mathcal{C}_i$, namely 
$$\mathbb{S}=\left\{\mathcal{S}\subseteq [n]: |\mathcal{S}|=k \text{ and } |\mathcal{S}\cap \mathcal{C}_i|\leq r \text{ for any } i\right\}.$$
Thus, establishing the lemma is equivalent to showing that the matrix $G_\mathcal{S}$ is invertible for $\mathcal{S}\in \mathbb{S}$. 
Recall that the permanent of any submatrix $r\times t$ of $A$ for $t\leq r$, is a monic polynomial in $\omega$, and $a$ is the largest degree of $\omega$  among the entries of $A$. Hence the determinant of $G_\mathcal{S}$ is a \emph{nonzero} polynomial in $\omega$  with coefficients in $\mathbb{F}_p$, and degree at most $ak$. However, since $\omega$  is a primitive element of the field $\mathbb{F}_{p^{ak+1}}$, the degree of its minimum polynomial in $\mathbb{F}_p[x]$ is \emph{exactly} $ak+1$, and the determinant is a nonzero element of $\mathbb{F}_{p^{ak+1}}$. Namely, $G_\mathcal{S}$ is invertible for any $\mathcal{S}\in \mathbb{S}$, and the result follows.

\end{IEEEproof}

We are now ready to prove the optimality of the code distance.
\begin{theo}
The minimum distance $d$ of the code generated by the matrix $G$ equals to
$$d=n-k-\left(\left(\left\lceil \frac{k}{r} \right\rceil -1\right)(\delta-1)+1\right)+2.$$
\end{theo}
\begin{IEEEproof}
We will prove the distance optimality  by showing that the value of the parameter $\mu$ in the matroid $\cM$ equals to $(\lceil k/r\rceil-1)(\delta-1)+1$, and then the result will follow from Proposition \ref{good thm}. 
Consider the first $r+\delta -1$ coordinates of the code generated by $G$. By construction these coordinates correspond to an $(r+\delta -1,r)$ MDS local code. Hence, in the matroid $\cM$ any $r+1$-subset of these coordinates forms a circuit. In particular for $i=1,\ldots,\delta-1$ the set $\mathcal{C}_i=[r]\cup \{r+i\}\subset [r+\delta-1]$ forms a circuit, furthermore this family of $\delta-1$ circuits has a nontrivial union of size $r+\delta -1$. 
In general, from each  MDS local code, one can find $\delta-1$ circuits in $\cM$ whose union is non trivial and is of size $r+\delta -1$. 
Notice that the MDS local codes ``live'' on disjoint coordinates, hence by considering $\lceil k/r\rceil-1$ distinct MDS local codes, one can find $(\lceil k/r\rceil-1)(\delta-1)$ circuits $\cC_i$ whose union is nontrivial and is of size at most
\begin{align}
\left|\bigcup_{i=1}^{(\lceil k/r\rceil-1)(\delta-1)}\cC_i\right|&=(\lceil k/r\rceil-1)(r+\delta-1)\nonumber\\
&=r\lceil k/r\rceil-r +(\lceil k/r\rceil-1)(\delta-1)\nonumber\\
&<k+(\lceil k/r\rceil-1)(\delta-1)\label{ujuj}.
\end{align}
We claim that the value of the parameter $\mu$ is greater than $(\lceil k/r\rceil-1)(\delta-1)$. Let $1\leq \gamma \leq (\lceil k/r\rceil-1)(\delta-1)$ be an integer, we will show that $\mu$ can not be equal to $\gamma$ by finding $\gamma$ circuits whose union is nontrivial, but is of size less than $k+\gamma$. Consider the first $\gamma$ circuits $\cC_i$, then
\begin{align}
k+(\lceil k/r\rceil-1)(\delta-1)&>\left|\bigcup_{i=1}^{(\lceil k/r\rceil-1)(\delta-1)}\cC_i\right|\nonumber\\
&=\left|\bigcup_{i=1}^{\gamma}\cC_i\right|+\sum_{i=\gamma+1}^{(\lceil k/r\rceil-1)(\delta-1)}\left|\cC_i\backslash \bigcup_{j=1}^{i-1}\cC_j\right|\nonumber\\
&\geq \left|\bigcup_{i=1}^{\gamma}\cC_i\right|+(\lceil k/r\rceil-1)(\delta-1)-\gamma \label{rmrm},
\end{align}
where the first inequality follows from \eqref{ujuj}, and \eqref{rmrm} follows since the union is nontrivial. Hence 
$$\left|\bigcup_{i=1}^{\gamma}\cC_i\right|<k+\gamma,$$
and $\mu\neq \gamma$.
We conclude that $\mu\geq (\lceil k/r\rceil-1)(\delta-1)+1$. 
Next we will show that any non trivial union of $(\lceil k/r\rceil-1)(\delta-1)+1$ circuits is of size at least $k+(\lceil k/r\rceil-1)(\delta-1)+1$, and hence $\mu=(\lceil k/r\rceil-1)(\delta-1)+1.$

Consider $(\lceil k/r\rceil-1)(\delta-1)+1$ circuits $\mathcal{C}_i$ whose union is non trivial. If at least one of the circuits is trivial, namely is of size $k+1$, then it is easy to see that the size of the union of the circuits $\cC_i$ is at least $k+(\lceil k/r\rceil-1)(\delta-1)+1$. If all the circuits are non trivial then by Lemma \ref{rfrf} each circuit $\mathcal{C}_i$ is contained in some MDS local code. In each MDS local code, one can find at most $\delta -1$ circuits whose union is non trivial. Therefore by the pigeonhole principle there are at least $\lceil k/r\rceil$ circuits from the circuits $\cC_i$, say $\mathcal{C}_1,\ldots,\mathcal{C}_{\lceil k/r\rceil}$, that belong to \emph{distinct} MDS local codes. Hence
\begin{align*}
\left|\bigcup_{i=1}^{(\lceil k/r\rceil-1)(\delta-1)+1}\mathcal{C}_i\right|&= \left|\bigcup_{i=1}^{\lceil k/r\rceil}\cC_i\right|+\sum_{i=\lceil k/r\rceil+1}^{(\lceil k/r\rceil-1)(\delta-1)+1}\left|\cC_i\backslash \bigcup_{j=1}^{i-1}\cC_j\right|\\
&\geq \left|\bigcup_{i=1}^{\lceil k/r\rceil}\mathcal{C}_i\right|+(\lceil k/r\rceil-1)(\delta-2)\\
&=\lceil k/r\rceil(r+1)+(\lceil k/r\rceil-1)(\delta-2)\\
&=\lceil k/r\rceil r+(\lceil k/r\rceil-1)(\delta-1)+1\\
&\geq k+(\lceil k/r\rceil-1)(\delta-1)+1,\\
\end{align*}
where the first inequality follows since the union is non trivial. Therefore $\mu=(\lceil k/r\rceil-1)(\delta-1)+1$ and the result on the minimum distance follows from Proposition \ref{good thm}.
\end{IEEEproof}

\section{Conclusions}

In this work we introduced a new family of optimal $(n,k,r)$ LRCs that are simple to construct.
The codes are based on re-encoding Reed-Solomon encoded symbols for the added property of locality.
To prove the optimality of code construction, we establish a connection between the minimum distance of the code and properties of the matroid represented by its generator matrix.
We concluded with a generalization of the construction to optimal $(n,k,r,\delta)$ LRCs.
Although our code constructions are simple, they require a large finite field.
 This, however, does not seem to be a significant practical problem since each field element requires $O(k \log n)$ bits to be represented.
Explicit constructions of optimal LRCs for the case when $r+1$ does not divide $n$ and for small finite fields remain as open problems.

\section{Acknowledgment}

 This research was supported in part by NSF grant CCF1217894. Moreover we would like to thank Uzi Tomo Magen for intriguing and useful discussions.

\bibliographystyle{ieeetr}

\bibliography{explicit_LRC_IT}

\end{document}